\documentclass[
    ,final            
  ]
  {aipproc}

\layoutstyle{8x11double}

\begin{document}

\title{The Simulation of the Inelastic Impact}

\author{Hiroto Kuninaka}{
  address={Graduate School of Human and Environmental Studies, 
Kyoto University, Sakyo-ku, Kyoto, Japan, 606-8501}
}

\author{Hisao Hayakawa}{
  address={Department of Physics, Yoshida-south campus, 
Kyoto University, Sakyo-ku, Kyoto, Japan, 606-8501}
}


\begin{abstract}
The coefficient of normal restitution (COR) in an oblique impact is
theoretically studied. 
Using a two-dimensional lattice models for an elastic disk 
and an elastic wall, 
we investigate the dependency of COR on an incident angle 
and demonstrate that COR can exceed one and have a peak 
against an incident angle in our simulation. 
Finally, we explain these phenomena based upon 
the phenomenological theory of elasticity.
\end{abstract}

\maketitle


\section{Introduction}
The coefficient of restitution (COR) $e$ 
is introduced to determine the post-collisional 
velocity in the normal collision of two materials 
and defined 
by ${\bf u}^{'} \cdot {\bf n}=-e {\bf u} \cdot {\bf n}$,
where ${\bf u}$ and ${\bf u}^{'}$ are respectively 
the velocity of the contact point of two colliding materials 
before and after the impact, and ${\bf n}$ is the normal unit vector 
of the tangential plane of them. 
In the oblique impact, it has become clear that COR depends on 
the incident angle as well as on the impact velocity. 
Louge and Adams\cite{louge} recently reported 
that COR increases as a linear function of 
the tangent of the incident angle in the oblique impact 
of a hard aluminum oxide sphere on a thick elastoplastic plate. 
They also suggested that COR can exceed $1$ in most grazing impacts.
In this proceeding, we carry out the two-dimensional simulation 
of the oblique impact 
and investigate the dependency of COR on the incident angle based upon 
the theory of elasticity.
\section{Model}
Our numerical model is a two-dimensional model which is composed of 
an elastic disk and an elastic wall\cite{kuninaka_jpsj2003}. 
Each of them is composed of 
randomly placed mass particles connected by nonlinear springs 
each other. Numbers of mass particles are $400$ for the disk 
and $2000$ for the wall. The width and the height of the wall are 
$8R$ and $2R$ respectively where $R$ is the radius of the disk. 
\begin{figure}[htp]
\includegraphics[width=0.35\textwidth]{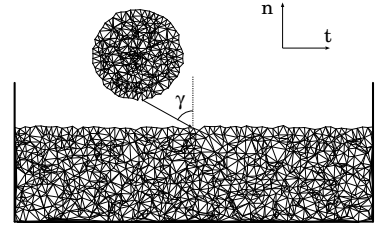}
\caption{The elastic disk and wall consisted of random lattice.}
\label{fig1}
\end{figure}
Both sides and bottom of the wall are fixed. 
Spring potential is described as 
$V(x)=\frac{1}{2} k_{a} x^{2}+\frac{1}{4} k_{b} x^{4}$,
where $x$ is a stretch from natural length. 
The values of $k_a$ are $1.0 \times m c^2/R^2$ for the disk 
and $k_a = 1.0 \times 10^{-2} m c^2/R^2$ for the wall, 
where $c$ is the one-dimensional sound of velocity. 
And we adopt $k_b = k_a \times 10^{-3}/R^2$ for each of them. 
Poisson's ratio $\nu$ can be evaluated from the strains of the band 
of random lattice 
in vertical and horizontal directions to the applied force. 
We obtain Poisson's ratio $\nu$ of the model as 
$\nu=(7.50 \pm 0.11) \times 10^{-2}$.

In our simulation, we define the incident angle $\gamma$ 
by the angle between the normal vector of the wall and the initial 
velocity vector of the disk(Fig.(\ref{fig1})). 
We fix the initial colliding velocity of the disk as $|{\bf v}(0)|=0.1c$ 
to control the normal and tangential components of 
the initial colliding velocity as 
$v_t(0)=|{\bf v}(0)|\sin\gamma$ and 
$v_n(0)=|{\bf v}(0)|\cos\gamma$, respectively. 
From the normal components of the contact point velocities 
before and after the collision, we calculate COR 
for each $\gamma$.
We use the fourth order symplectic numerical method 
for the numerical scheme of integration 
with the time step $\Delta t = 10^{-3}R/c$.
\section{Results and Discussion}
\begin{figure}[htbp]
\includegraphics[width=0.35\textwidth]{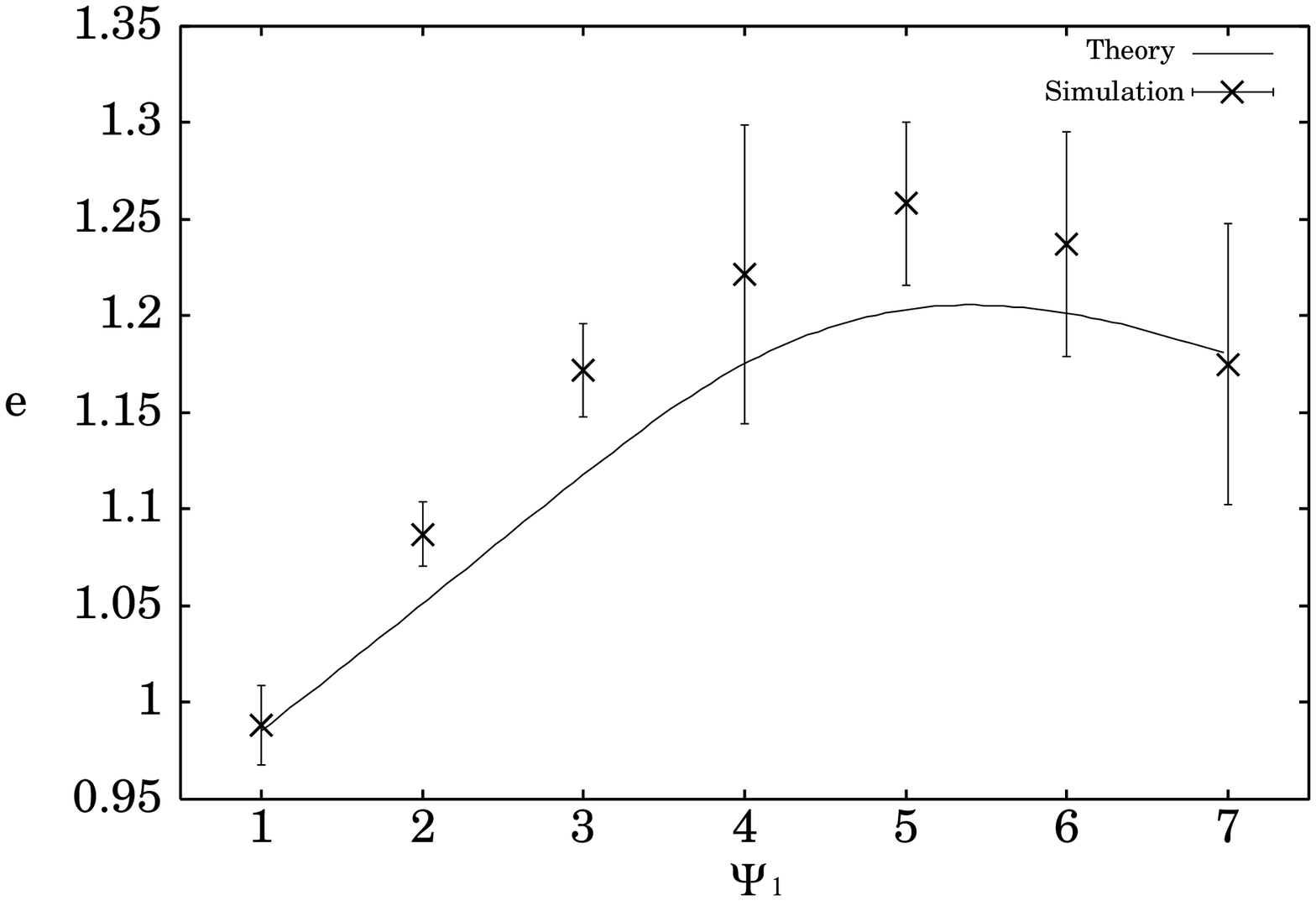}
\caption{Numerical and theoretical results of the relation between
 $\Psi_1$ and COR.}
\label{fig2}
\end{figure}

Figure \ref{fig2} is the COR against the tangent of 
the incident angle $\Psi_1=\tan\gamma$ in our simulation. 
The cross points are the average and the error bars are  
the standard deviation of 100 samples for each incident angle. 
This result shows that the COR increases as $\Psi_1$ increases 
to exceed $1$, and has a peak around $\Psi_1 = 5.0$.
This behavior is contrast to that in the experiment 
by Louge and Adams\cite{louge}.

\begin{figure}[htbp]
\includegraphics[width=0.35\textwidth]{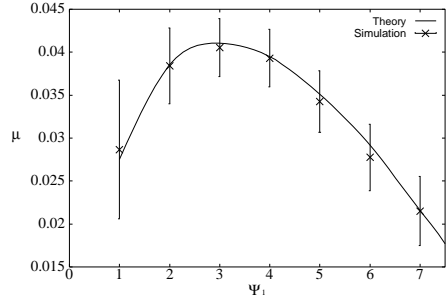}
\caption{Numerical and theoretical results of the relation between
 $\Psi_1$ and $\mu$.}
\label{fig4}
\end{figure}

To explain this result, we consider the correction of COR 
by the local deformation of the wall. 
We assume that the normal unit vector ${\bf n}$ to the surface of the wall 
rotates toward the incoming disk by an angle $\alpha$ to become 
${\bf n}^{\alpha}$. 
If we define $e = -({\bf u}^{'}\cdot{\bf n})/
({\bf u}\cdot{\bf n})$ 
and 
$e^{\alpha} = -({\bf u}^{'}\cdot{\bf n}^{\alpha})/
({\bf u}\cdot{\bf n}^{\alpha})$, 
the relation between 
$e$ and $e^{\alpha}$ becomes
\begin{equation} \label{COR}
e=(e^{\alpha}+\Psi_2^{\alpha} \tan\alpha)/(1-\Psi_1^{\alpha}
\tan\alpha),
\end{equation}
where $\Psi_{1}^{\alpha}=(\Psi_1-\tan\alpha)/
(1+\Psi_1 \tan\alpha)$ and 
$\Psi_2^{\alpha}=(\Psi_1-\tan\alpha)/
(1+\Psi_1 \tan\alpha)-3(1+e^{\alpha})
(\mu+\tan\alpha)/(1-\mu\tan\alpha)$.  
As for $\Psi_2^{\alpha}$, we use the phenomenological theory 
for the oblique impact by Walton and Braun\cite{walton}. 
The correction angle $\alpha$ can be estimated by the theory 
of elasticity. If we express the contact area by a parabora, 
$\tan\alpha$ equals to 
$|x_c-x_a|(1-2\theta)/R(2-2\theta)$ with 
$\theta = (1/\pi)\arctan (1-2\nu)/(\mu (2-2\nu))$, 
where $\mu$ is the coefficient of friction 
and $x_c$ and $x_a$ are the x coordinates of both ends of the contact
area. $\mu$ can be calculated from the simulation data through the definition 
$\mu=|J_t|/|J_n|$, where $J_n$ and $J_t$ are the normal and tangential 
components of the impulse. The relation between $\mu$ and $\Psi_1$ are 
cross points in Fig. \ref{fig4}. 
The solid curve in Fig. \ref{fig2}  is Eq. (\ref{COR}) with 
$e^{\alpha}=0.95$ which is COR in the normal impact. Our numerical
results can be reproduced by our phenomenological theory. 

The relation between $\mu$ and $\Psi_1$ can be explained as follows. 
We assume that jags are uniformly placed on the surface of the
wall with the density $\rho$ per unit length and 
the tangential velocity of the disk 
is decreased by $\eta$ when the disk interacts with the one jag.  
The tangential and normal impulses can be calculated by calculating  
the number of jags the disk interacts during collision time\cite{ces} 
as $J_t =-m\eta\rho |{\bf v}(0)|\sin\gamma \pi (R/c)
\sqrt{\ln(4 c/|{\bf v}(0)|\cos\gamma)}$ and 
$J_n = -m(e+1) |{\bf v}(0)|\cos\gamma$. 
We also assume that the tangential impulse decreases by 
$J_t^{'}=-m\zeta |{\bf v}(0)| \sin\gamma$ which is proportional to 
the initial tangential velocity with the proportionality constant
$\zeta$. Thus, $\mu$ can be calculated by the ratio of $|J_t-J_t^{'}|$ 
to $J_n$ as 
\begin{equation}\label{toy}
\mu =\left|\zeta\tan\gamma-\eta\rho\tan\gamma\pi \frac{R}{c}
\sqrt{\ln\left(\frac{4 c}{|{\bf v}(0)|\cos\gamma}\right)}\right|
/(e+1).
\end{equation}
The solid curve in Fig.\ref{fig4} is Eq.(\ref{toy}) 
with  $\zeta=0.317$ and $\eta\rho=0.0416c/R$. 
Our numerical results can be well reproduced by our phenomenological 
theory of the coefficient of friction.

\section{Summary}
In the present study, we have shown that COR can exceed $1$ also 
in our two-dimensional simulation and depends on $\mu$ 
in the oblique impact. Our results can be explained by  
our simple phenomenological theory.


\begin{theacknowledgments}
This study is partially supported by the Grant-in-Aid of 
Ministry of Education, Science and Culture, Japan (Grant No. 15540393).
\end{theacknowledgments}


\bibliographystyle{aipproc}   

\bibliography{impact}

\IfFileExists{\jobname.bbl}{}
 {\typeout{}
  \typeout{******************************************}
  \typeout{** Please run "bibtex \jobname" to optain}
  \typeout{** the bibliography and then re-run LaTeX}
  \typeout{** twice to fix the references!}
  \typeout{******************************************}
  \typeout{}
 }

\end{document}